\documentstyle[aps,amsfonts]{revtex} 
\begin{document} 
 
\twocolumn[\hsize\textwidth\columnwidth\hsize\csname 
@twocolumnfalse\endcsname 
 
\date{September 29, 1999} 
\title{Domain walls junctions from a distance.} 
\author{Daniele Binosi} 
\address{ 
Dep.to de Fis\`\i ca Teor\`\i ca and IFIC, Centro Mixto, Universidad de
Valencia-CSIC, Valencia, Spain\\ 
$\mathrm e$-$\mathrm mail: binosi@titan.ific.uv.es$} 
\maketitle 
\begin{abstract} 
It is pointed out that it is 
very unlikely that we can find analytical solutions built up from
elementary functions 
for the domain wall junctions problem in Wess-Zumino models possessing a
trivial K\"ahler potential.
\end{abstract} 
\vskip2pc] 
 
Recently BPS saturated intersecting domain walls (BPS junctions from now
on)  
have attracted attention of many authors  
\cite{gibbons,trodden,shifman}. In particular, it has been argued  
\cite{gibbons,trodden} that in $N=1$ supersymmetric theories BPS junctions  
preserves $1/4$ of the original supersymmetry, {\it i.e.} only one  
super charge (out of four) acts trivially on the junctions. 

This fact  
makes particularly interesting the question of whether or not such
junctions (if  
they exist at all) can be constructed explicitly. 
 
In \cite{trodden} this problem was studied in detail; in Sec. VI of this  
paper an explicit solution was suggested in the case of a generalized
Wess-Zumino
model possessing a ${\mathbb Z}^4$ symmetry. Since the Ansatz proposed
there 
for the superpotential profile ${\mathcal W}(z,\bar z)$
seems the only reasonable one we can write in terms of elementary
function, we can address the issue of its 
validity in the case of trivial K\"ahler potential (in \cite{trodden} it
was shown that this is a solution of the
BPS equations for some non-trivial K\"ahler potential). 
 
To be specific, we consider a Wess-Zumino model describing a single chiral  
superfield with the superpotential 
\begin{eqnarray} 
{\mathcal W}=\Phi-\frac1{N+1}\Phi^{N+1}. 
\label{superp} 
\end{eqnarray}
 
This model exhibits a ${\mathbb Z}^N$ symmetry, with $N$ chirally
asymmetric  
vacuum states at   
\begin{eqnarray}  
\phi_k=\omega^k, \quad k=0,1,\dots,N-1, \quad  
\omega=\exp\left(\frac{2\pi i}N\right), 
\end{eqnarray}
where $\phi$ represents the lowest component of the superfield $\Phi$.
 
We are interested in the BPS junctions interpolating amongst these vacua,
and  
which satisfy the creek equation \cite{gibbons,shifman1}
\begin{eqnarray} 
\partial_z\phi=\frac12\left(1-\bar\phi^N\right), 
\label{creek} 
\end{eqnarray} 
where the usual phase appearing in the r.h.s. of (\ref{creek}) 
has been absorbed via a suitable redefinition of the complex coordinate
$z=x+iy$.

Note, in fact, that unlike the case of the domain walls one has to promote
$z$ to be  
a complex variable; moreover no holomorphic dependence for the field  
$\phi$ on $z$ is allowed (in such a case we can basically patch together 
domain wall solution in each sector, via a conformal mapping such as
$z\to\widetilde z^N$: this
is anyway not what we would like to call a junction). 
 
It is useful then to introduce the parameterization
$\phi=\exp(\varepsilon+i\sigma)$ and 
$z=r\exp(i\theta)$, so that 
\begin{eqnarray} 
2\partial_z=\exp(-i\theta)\left(\partial_r-\frac ir\partial_\theta\right). 
\end{eqnarray} 
Equation (\ref{creek}) then ``reduces'' to the following system of  
{\it coupled non linear partial differential equations}: 
 
\begin{eqnarray} 
\partial_r\varepsilon+\frac1r\partial_\theta\sigma & = &  
\exp(-\varepsilon)\cos(\theta-\sigma) \nonumber \\ 
& & -\exp[(N-1)\varepsilon]\cos(\theta-\sigma) 
\cos(N\sigma)  \nonumber \\ 
& & -\exp[(N-1)\varepsilon]\sin(\theta-\sigma)\sin(N\sigma),  
\label{system1}\\ 
\partial_r\sigma-\frac1r\partial_\theta\varepsilon & = &  
\exp(-\varepsilon)\sin(\theta-\sigma) \nonumber \\ 
&  & +\exp[(N-1)\varepsilon]\cos(\theta-\sigma) 
\sin(N\sigma) \nonumber \\ 
& & -\exp[(N-1)\varepsilon]\sin(\theta-\sigma)\cos(N\sigma). 
\label{system2} 
\end{eqnarray} 

Trying to find analytical solutions of this system is out of question.
However, we can look at its 
asymptotics, by studying the behavior of the field profile $\phi$ in the
sector $0<\theta<2\pi/N$ 
when $r\to\infty$. In this sector at $r\to\infty$ one has
$\varepsilon,\sigma\to0$, so that, 
linearizing the previous equations, we find 
\begin{eqnarray} 
\partial_r\varepsilon+\frac1r\partial_\theta\sigma & = &
-N\varepsilon\cos\theta-N\sigma\sin\theta 
\label{linear1}\\ 
\partial_r\sigma-\frac1r\partial_\theta\varepsilon & = &
-N\varepsilon\sin\theta+N\sigma\cos\theta 
\label{linear2} 
\end{eqnarray} 
 
Despite of its cute form (it actually says to us that the differential
operator appearing 
in the l.h.s. of (\ref{linear1}) and (\ref{linear2}) acts on the solution
$(\varepsilon,-\sigma)$ 
as a rotation) we were unable to find the general solution of this system. 
 
Let us consider the case $N=4$. In this case 
the following Ansatz for the superpotential profile has been proposed in
\cite{trodden} 
\begin{eqnarray} 
{\mathcal W}(z,\bar z) & = & \left(\frac{2-2i}5\right)\left\{\tanh(z+\bar
z)-i\tanh[i(z-\bar z)]\right\} \nonumber \\ 
& = &
\left(\frac{2-2i}5\right)\left[\tanh(2r\cos\theta)+i\tanh(2r\sin\theta)\right],
\nonumber \\ 
\label{Ansatz} 
\end{eqnarray} 
(our $\mathcal{W}$ is their $W$, and we set for convenience $\Lambda=1$). 
 
Despite the fact that Eq. (\ref{Ansatz}) represents the superpotential
profile rather then the junction profile,  
we can extract from it important information, as the asymptotic behavior
of the junction.
 
Recall in fact that we are considering  the sector $0<\theta<\pi/2$ and
$r\to\infty$, 
where the field must approach the vacuum value $\phi_0=1$ (and ${\mathcal
W}(\phi_0)=1$). 
First of all note that in this region we have
\begin{eqnarray} 
\tanh(2r\cos\theta)\sim 1-\alpha, & \quad & \alpha=2\exp[-4r\cos\theta]>0,
\\ 
\tanh(2r\sin\theta)\sim 1-\beta, & \quad & \beta=2\exp[-4r\sin\theta]>0,  
\end{eqnarray}  
so that 
\begin{eqnarray} 
{\mathcal W}(r,\theta)\sim
\frac45-\frac25(\alpha+\beta)+\frac25i(\alpha-\beta).  
\label{1} 
\end{eqnarray} 
On the other hand, since 
\begin{eqnarray} 
\phi=\exp(\varepsilon+i\sigma)\sim 1+\varepsilon+i\sigma,
\end{eqnarray}
we also get 
\begin{eqnarray} 
{\mathcal W}=\phi-\frac15\phi^5\sim\frac45-2(\varepsilon+i\sigma)^2. 
\end{eqnarray} 

In this way, we get the following system of equations for 
$\varepsilon$ and $\sigma$ 
\begin{eqnarray} 
& & \varepsilon^2-\sigma^2 =\frac15(\alpha+\beta), \label{sl1} \\ 
& & \varepsilon\sigma = \frac1{10}(\beta-\alpha),  
\label{sl2} 
\end{eqnarray} 
which has the solutions 
\begin{eqnarray} 
\sqrt{10}\varepsilon & = & \pm
\sqrt{(\alpha+\beta)+\sqrt{2(\alpha^2+\beta^2)}}, \\ 
\sqrt{10}\sigma & = & \pm
{\sqrt{-(\alpha+\beta)+\sqrt{2(\alpha^2+\beta^2)}}}, 
\end{eqnarray} 
(due to the fact that $\alpha,\ \beta>0$ and that $\varepsilon$ and
$\sigma$ are real,  
two solutions of the quartic equation coming from the system above can be
discarded). 
 
Now, both $\varepsilon$ and $\sigma$ are of order
$\sqrt\alpha\sim\sqrt\beta\sim\exp(-2r)$; 
thus we must check that at  
this order they satisfy the linearized equation of motion. 
Let us look at the + solution (similar results will hold true for the $-$
solution);  
in this case, supposing $\alpha\neq\beta$, 
we get the following relation for Eq. (\ref{linear1}) (recall that $N=4$) 
\begin{eqnarray} 
& & \left\{\cos\theta\left[-\alpha\sigma\left(1+
\frac{2\alpha}{\sqrt{2(\alpha^2+\beta^2)}}\right)+\frac15\left(\beta-\alpha\right)\varepsilon
\right. \right.  
\nonumber \\
& & \left.
-\beta\varepsilon\left(-1+\frac{2\beta}{\sqrt{2(\alpha^2+\beta^2)}}\right)\right]
\nonumber \\
& &
+\sin\theta\left[-\beta\sigma\left(1+\frac{2\beta}{\sqrt{2(\alpha^2+\beta^2)}}\right)+
\frac15\left(\beta-\alpha\right)\sigma \right. \nonumber \\
& &
\left.\left.+\alpha\varepsilon\left(-1+\frac{2\alpha}{\sqrt{2(\alpha^2+\beta^2)}}\right)\right]\right\}=0.
\label{eq}
\end{eqnarray} 
 
Now each square bracket must be equal to zero, and it is easy to see that
this is not the case.  

In the case $\alpha=\beta$, {\it i.e.} on the line of constant phase
$\theta=\pi/4$, Eq. (\ref{eq}) gets
simplified (in particular note that in this case $\sigma=0$) but the same
result holds: it is not satisfied.
 
Thus the superpotential profile (\ref{Ansatz}), 
despite the fact that it has the correct asymptotic behavior 
(reproducing correctly all the values of the superpotential in the vacua
of the theory) and satisfies 
the integral of motion $\partial_x{\mathcal W}_2=\partial_y{\mathcal W}_1$
(where we have switched to 
the parameterization ${\mathcal W}={\mathcal W}_1+i{\mathcal W}_2$ and
$z=x+iy$), does not go trough the BPS
equations, in the case of trivial K\"ahler potential.

Following these lines of reasoning we also tried to see if it was possible
to generalize this Ansatz, for 
example by choosing as arguments of the $\tanh$'s in (\ref{Ansatz}) 
two arbitrary functions $f$ and $g$ with appropriate dependence 
on $z$ and $\bar z$ and suitable boundary conditions in each different
sector of the $N=4$ junction.  
Unfortunately this generalized ansatz for the superpotential profile 
${\mathcal W}(z,\bar z)$ fails too: 
finding $f$ and $g$ is of the same difficulty of finding the solution of
our original creek equation. 
Moreover, such an Ansatz for the junction profile $\phi(z,\bar z)$ fails
as well: it simply does not go
through the linearized equations of motion. 
 
Concluding, given the complexity of the creek equation (\ref{creek}), and
the fact that one is able to write
down an Ansatz for the profile functions only in the case $N=4$ (as first
done in \cite{trodden}), 
it seems to us very unlikely that we will find analytical solutions
describing BPS junctions (or even only 
their asymptotic behavior) in Wess-Zumino models with trivial K\"ahler
potential; 
numerical investigation is the only way to shed light on this 
issue.\\

Useful discussion with I. Kogan, M. Shifman, A. Ritz and V. Vento are
gratefully acknowledged. 
We acknowledge warm hospitality extended to us at the Theoretical Physics
Institute 
of the University of Minnesota. This work has been partially supported by
the DGICYT-PB97-1127 
grant.

\end{document}